\documentclass[aps,pra,10pt,twocolumn,superscriptaddress]{revtex4-2}
\usepackage{CJK}
\usepackage[T1]{fontenc}    
\usepackage[utf8]{inputenc} 
\usepackage{lmodern}        
\usepackage{graphicx}
\usepackage{dcolumn}
\usepackage{bm}
\usepackage{hyphenat} 
\usepackage{physics}
\usepackage{mathrsfs}
\usepackage{amsmath}

\usepackage{float}
\usepackage{fancyhdr}
\usepackage{fnpos}
\usepackage[english]{babel}
\usepackage{mathptmx}
\usepackage[bookmarks=false,bookmarksopen=false,pdfpagelayout=SinglePage,pdfstartview=Fit]{hyperref}
\hypersetup{colorlinks,breaklinks,linkcolor=blue,menucolor=green,urlcolor=blue,citecolor=blue}
\usepackage[dvipsnames]{xcolor}

\usepackage{epstopdf}
\newcommand{\threej}[6]{%
	\begin{pmatrix}
		#1 & #2 & #3 \\
		#4 & #5 & #6
	\end{pmatrix}%
}	
\begin{document}
	
\title{Perturbation-assisted Observation of the Lowest Vibrational Level of the $\mathrm{b}^{3}\Pi_{0}$ State of Ultracold LiK Molecules}
\author{Anbang~Yang}
\thanks{These authors contributed equally.}
\affiliation{Centre for Quantum Technologies (CQT), 3 Science Drive 2, Singapore 117543}
\author{Xiaoyu~Nie}
\thanks{These authors contributed equally.}
\affiliation{Centre for Quantum Technologies (CQT), 3 Science Drive 2, Singapore 117543}
\author{Hao~Lin~Yu}
\affiliation{Centre for Quantum Technologies (CQT), 3 Science Drive 2, Singapore 117543}
\author{Yiming~Liu}
\affiliation{Centre for Quantum Technologies (CQT), 3 Science Drive 2, Singapore 117543}
\author{Victor~Avalos}
\affiliation{Centre for Quantum Technologies (CQT), 3 Science Drive 2, Singapore 117543}
\author{Canming~He}
\affiliation{Department of Physics, National University of Singapore, 2 Science Drive 3, Singapore 117542}
\author{Jacek~Klos}
\affiliation{Department of Physics, Temple University, Philadelphia, PA, 19122, USA}
\author{Svetlana~Kotochigova}
\affiliation{Department of Physics, Temple University, Philadelphia, PA, 19122, USA}
\author{Kai~Dieckmann}
\email[Electronic address:]{phydk@nus.edu.sg}
\affiliation{Centre for Quantum Technologies (CQT), 3 Science Drive 2, Singapore 117543}
\affiliation{Department of Physics, National University of Singapore, 2 Science Drive 3, Singapore 117542}

{\tiny }

\date{\today}

\begin{abstract}
The narrow transition from the lowest rovibrational level of the $\mathrm{X}^{1}\Sigma^{+}$ electronic ground state to the lowest vibrational level of the $\mathrm{b}^{3}\Pi_{0}$ potential provides opportunities for achieving magic-wavelength trapping of ultracold bialkali molecules for enhancing their rotational coherence times. Guided by existing spectroscopic data of several perturbed and deeply-bound rovibrational states of the $\mathrm{A}^{1}\Sigma^{+}$ potential [Grochola et al., \textit{Chem. Phys. Lett.}, 2012, \textbf{535}, 17-20], we conducted a targeted spectroscopic search and report the first observation of the lowest vibrational level of the $\mathrm{b}^{3}\Pi_{0}$ state in $^{6}\mathrm{Li}^{40}\mathrm{K}$. The transition frequency from $\ket{\mathrm{X}^{1}\Sigma^{+},\,v=0,\,J=0}$ to $\ket{\mathrm{b}^{3}\Pi_{0},\,v'=0,\,J'=1}$ is determined to be 314,230.5(5) GHz. Assisted by microwave spectroscopy, we resolved the rotational structure of $\ket{\mathrm{b}^{3}\Pi_{0},\,v'=0}$ and extracted a rotational constant of $h\times8.576(44)$ GHz for the $\mathrm{b}^{3}\Pi_{0}$ state. From this, we deduced an energy separation between $\ket{\mathrm{b}^{3}\Pi_{0},\,v'=0,\,J'=0}$ and $\ket{\mathrm{X}^{1}\Sigma^{+},\,v=0,\,J=0}$ of $hc\times$10,481.03(2) $\mathrm{cm}^{-1}$. Our work provides timely and precise information on the deeply-bound region of the $\mathrm{b}^{3}\Pi_{0}$ triplet excited potential of LiK, and benefits future applications of ultracold LiK isotopologues in quantum simulation and quantum computation that demand long coherence times.
\end{abstract}

\maketitle

\section{Introduction}
The lowest vibrational level $v'=0$ of the $\mathrm{b}^{3}\Pi_{0}$ potential\footnote{As the $\mathrm{b}^{3}\Pi_{0}^{-}$ state does not optically couple to $\mathrm{X}^{1}\Sigma^{+}$, for the rest of the article the $\mathrm{b}^{3}\Pi_{0}^{+}$ state is abbreviated as $\mathrm{b}^{3}\Pi_{0}$} of bialkali molecules is perturbed by the $\mathrm{A}^{1}\Sigma^{+}$ electronic potential due to spin-orbit coupling (SOC). Therefore, it weakly couples to the rovibrational ground state of the $\mathrm{X}^{1}\Sigma^{+}$ potential via an electronic dipole transition\cite{Vexiau2017}. As the equilibrium distance of $\mathrm{b}^{3}\Pi_{0}$ state matches well with $\mathrm{X}^{1}\Sigma^{+}$, the $\ket{\mathrm{b}^{3}\Pi_{0}, v'=0}$ state decays mainly to $\ket{\mathrm{X}^{1}\Sigma^{+},v=0}$, which indicates for a long excited-state lifetime and a small scattering rate $\Gamma$. We can tune a laser near the transition frequency from the $\ket{\mathrm{X}^{1}\Sigma^{+}, v=0}$ state to $\ket{\mathrm{b}^{3}\Pi_{0}, v'=0}$, such that the polarizabilities of the two lowest rotational levels in the $\ket{\mathrm{X}^{1}\Sigma^{+}, v=0}$ manifold become equal, or one of them vanishes. These conditions are referred to as the \textit{magic} or \textit{tune-out} conditions \cite{Bause2020}. Coherence times over 1 second of the microwave transition between the two lowest rotational level of the $\ket{\mathrm{X}^{1}\Sigma^{+}, v=0}$ state have been observed under magic conditions \cite{Gregory2024}. Such long coherence times enable the exploration of quantum many-body physics with long-range dipole-dipole interactions\cite{Carroll2025}, studying SU(\textit{N}) physics \cite{Mukherjee2025}, and performing quantum information processing with ultracold polar molecules \cite{Ruttley2024,Ruttley2025,Picard2025}.

Systematic experimental studies of the $\mathrm{b}^{3}\Pi_{0}$ potential for bialkali molecules and determination of the frequency for the transition from $\ket{\mathrm{b}^{3}\Pi_{0},v'=0}$ to the absolute ground state are limited to heavy molecules \cite{Bause2020,Kobayashi2014,Harker2015}. So far, the spectroscopic study of LiK molecules has focused on singlet potentials \cite{Bednarska1997,Pashov1998, Rousseau1999,Ridinger2011,Grochola2012,Botsi2022}, as they strongly couple to the singlet $\mathrm{X}^{1}\Sigma^{+}$ ground state. Here, we report the first observation of the lowest vibrational state of the $\mathrm{b}^{3}\Pi_{0}$ potential for LiK molecules. To the best of our knowledge, there are no experimental studies of deeply bound molecular states in excited triplet potentials of LiK. The transition frequency is first predicted by ab initio electronic potentials for LiK and then corrected by our analysis of the discarded data of perturbed lines observed in the $\mathrm{A}^{1}\Sigma^{+}$ state\cite{Allouche,Grochola2012}. For probing the transition from $\ket{\mathrm{X}^{1}\Sigma^{+}, v=0, J=0}$ to $\ket{\mathrm{b}^{3}\mathrm{\Pi}_{0}, v'=0, J'=1}$, we prepared a tunable diode laser system at 954 nm. When the laser is tuned near the resonance, the ac-Stark shift induced by the laser leads to a two-photon detuning for the stimulated Raman adiabatic passage (STIRAP) process for creating ultracold ground state $^{6}\mathrm{Li}^{40}\mathrm{K}$ molecules and causes molecule losses. The resonance frequency is determined to be 314.2305(5) THz at the center of the observed loss feature. 
\begin{figure}[t]
	\includegraphics[width = 8.6cm]{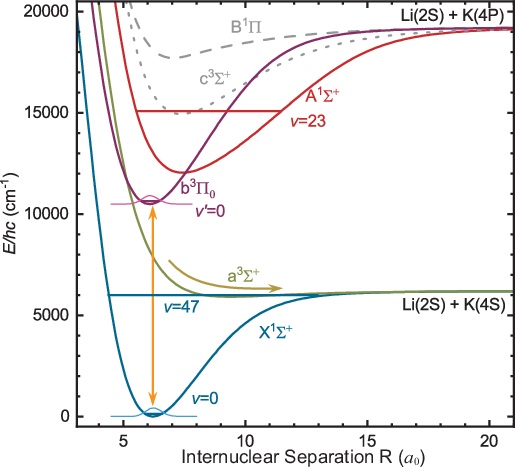}
	\centering
	\caption{Non-relativistic electronic potential energy curves for LiK\cite{Allouche} as functions of internuclear separation $R$. $\mathrm{X}^{1}\Sigma^{+}$ (blue curve) and $\mathrm{a}^{3}\Sigma^{+}$ (green curve) ground state potentials at large $R$ connect to the Li(2S) plus K(4S) atomic asymptote. Two relevant excited electronic states are the $\mathrm{A}^{1}\Sigma^{+}$ (red curve) and $\mathrm{b}^{3}\Pi_{0}$ (purple curve), which are coupled by the relativistic spin-orbit interaction. The orange arrow indicates the transition from $\ket{\mathrm{X}^{1}\Sigma^{+},v=0}$ to $\ket{\mathrm{b}^{3}\Pi_{0},v'=0}$. $\ket{\mathrm{b}^{3}\Pi_{0},v'=0}$ can also decay via pre-dissociation process due to coupling to the $\mathrm{a}^{3}\Sigma^{+}$ state as indicated by the curved arrow. The $\mathrm{B}^{1}\Pi$ (gray dashed line) and $\mathrm{c}^{3}\Sigma^{+}$ (gray dotted line) excited states are not considered in this work.}
	\label{fig:PECs}
\end{figure}
\section{Ultracold $^{6}\mathrm{Li}^{40}\mathrm{K}$ Molecules}
\subsection{Production of molecules}
\label{sec:description}
In our experiment, around $3,000$-$10,000$ $^{6}\mathrm{Li}^{40}\mathrm{K}$ molecules are produced for each cycle. The experimental sequence starts by the magneto-optical trapping of $^{87}\mathrm{Rb}$, $^{6}\mathrm{Li}$ and $^{40}\mathrm{K}$ atoms simultaneously inside an ultrahigh-vacuum stainless-steel chamber\cite{Taglieber2006}. After optical pumping, these spin-polarized atoms are transferred to a quartz cell with improved vacuum conditions and high optical access. Evaporative cooling of trapped Rb atoms in a magnetic trap in the quadrupole-Ioffe configuration is performed by driving the hyperfine microwave transitions around 6.8 GHz, which allows for the sympathetic cooling of fermionic Li and K atoms. At the end of evaporation quantum degenerate gases of lithium and potassium are obtained, which are transferred to a crossed optical dipole trap. Weakly-bound Feshbach molecules are created via magneto-association in the ultracold atomic mixture of approximately $5\times10^4$ $^{6}\mathrm{Li}$ and $1\times10^{5}$ $^{40}\mathrm{K}$ atoms by adiabatically ramping down the magnetic bias field across the Feshbach resonance at 21.56 mT \cite{Wille2008,Tiecke2010,Taglieber2008b,Voigt2009} with a total magnetic quantum number of $M=-5$. Such Feshbach molecules have a strong spin-singlet character, which is contributed by the $\ket{\mathrm{X}^{1}\Sigma^{+},v=47}$ state of $^{6}\mathrm{Li}^{40}\mathrm{K}$. These molecules are transferred to the singlet rovibrational ground state $\ket{\mathrm{X}^{1}\Sigma^{+}, v=0, J=0}$ via the singlet excited state $\ket{\mathrm{A}^{1}\Sigma^{+},v'=23,J'=1}$ by STIRAP with a one-way efficiency of 96\% \cite{Yang2020,He2024}. The relevant states are labeled in Fig.~\ref{fig:PECs}. After spectroscopy pulses, the ground-state molecules are transferred back to the Feshbach state for detection.
\subsection{Potential energy curves}
\label{sec:PEC}
We use the ab initio non-relativistic electronic potentials computed by Allouche et al.\cite{Allouche, Rousseau1999}. The molecular constants obtained from these potential energy curves are consistent with the ones obtained from recently published LiK molecular potentials\cite{Miadowicz2013,Musial2016} for the low-lying $^{1}\Sigma^{+}$, $^{3}\Sigma^{+}$ and $^{1}\Pi$ states. For these states the typical discrepancies from experimental values are a few percent. The eigen-energies and wavefunctions of the vibrational levels can be obtained by solving the Schr\"{o}dinger equation with the potentials using the Fourier-grid Hamiltonian (FGH) method\cite{Marston1989}. As shown in Fig.~\ref{fig:PECs} the equilibrium position $R_{e}^{b} = 6.08\,\,a_{0}$ of the $\mathrm{b}^{3}\Pi_{0}$ potential matches well with $R_{e}^{X}=6.24\,\,a_{0}$ of the $\mathrm{X}^{1}\Sigma^{+}$ potential. $a_{0}$ is the Bohr radius. This leads to a Franck-Condon factor (FCF) of 0.90, which is the square of the wavefunction overlapping integral between the lowest vibrational states of the two potentials.
\section{Spectroscopy}
\subsection{Estimation of transition strength}
\label{sec:transitionstrength}
\begin{figure}[t]
	\centering
	\includegraphics[]{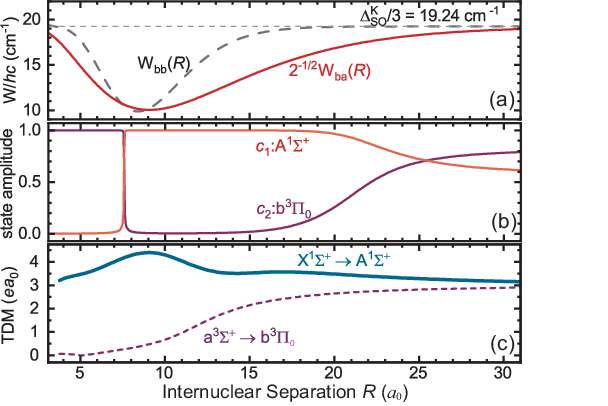}
	\caption{\label{fig:SOCs}Spin-orbit matrix elements (panel a), state amplitudes $c_1$ and $c_2$ for states A$^1\Sigma^+$ and b$^3\Pi$, respectively (panel b), and electric transition dipole moments (panel c) as functions of internuclear separation $R$. In panel (a) quantity $\Delta^{\rm K}_{\rm{SO}}$ is the spin-orbit splitting between the 4p(P$_{1/2}$) and 4p(P$_{3/2}$) levels of the potassium atom. In panel (c) the dipole moments are in atomic units $ea_0$.}
\end{figure}

The deeply-bound rovibrational states of the triplet $\mathrm{\mathrm{b}^{3}}\Pi_{0}$ potential can carry spin-singlet characters caused by perturbations from nearby $\mathrm{A}^{1}\Sigma^{+}$ states via spin-orbit interactions. In such cases, the wavefunction of a perturbed $\mathrm{b}^{3}\Pi_{0}$ state $\ket{\Psi^{b}_{v',\,J',\,m_{J'}}}$ can be expressed as:
\begin{equation}
	\ket{\Psi^{b}_{v',\,J',\,m_{J'}}} = ( c_{1}\ket{\mathrm{A}^{1}\Sigma^{+}}+c_{2}\ket{\mathrm{b}^{3}\Pi_{0}})\otimes\ket{\psi^{b}_{v'}}\otimes\ket{\Phi_{J',\,m_{J'}}}.
\end{equation}
Here, $c_{1}$ and $c_{2}$ are the amplitudes of the electronic wavefunction on the states $\ket{\mathrm{\mathrm{A}^{1}\Sigma^{+}}}$ and $\ket{\mathrm{\mathrm{b}^{3}\Pi_{0}}}$. $\ket{\psi^{b}_{v'}}$ and $\ket{\Phi_{J',\,m_{J'}}}$ represents the vibrational and rotational wavefunctions respectively. The ground state is purely spin-singlet and is therefore expressed as $\ket{\mathrm{X}^{1}\Sigma^{+},v=0,J}$. The values of $c_{1}$ and $c_{2}$, as shown in Fig.~\ref{fig:SOCs}(b), can be calculated by diagonalizing the spin-orbit interaction Hamiltonian matrix plus the energies of the states in the uncoupled state basis. The diagonal and off-diagonal elements of the SOC matrix are shown in Fig.~\ref{fig:SOCs} (a). $c_{1}$ is estimated to be 0.0062 for $\ket{\mathrm{b}^{3}\Pi_{0},v'=0}$ using $R_{0}^{b}=6.08\,\mathrm{a}_{0}$, as the ground state wavefunction is Gaussian-like. This factor strongly depends on the position where the A and b potential curves cross and on the magnitude of the SOC elements. $\ket{\psi^{b}_{v'}}$ and $\ket{\Phi_{J',\,m_{J'}}}$ are the vibrational and rotational wavefunctions.
The transition from $\ket{\mathrm{X}^{1}\Sigma^{+},v=0,J=0}$ to $\ket{\mathrm{b}^{3}\Pi_{0},v'=0,J'=1}$ is therefore an electronic dipole transition, and the corresponding transition dipole matrix element $d_{i,j}$ can be calculated by 
\begin{align*}
	d_{i,j}&=\bra{\Psi^{b}_{v',\,J',\,m_{J'}}}\hat{d}\ket{\mathrm{X}^{1}\Sigma^{+},v,J,m_{J}}\\
&	=\sqrt{(2J'+1)(2J+1)}\threej{J'}{1}{J}{-m_{J}'}{p}{m_{J}}\threej{J'}{1}{J}{0}{0}{0}\\ &\hspace{0.5cm}\times c_{1}\bra{\mathrm{A}^{1}\Sigma^{+}}d_{0}\ket{\mathrm{X}^{1}\Sigma^{+}}_{R=R_{\mathrm{C}}}\bra{\psi_{v'}^{b}}\ket{\psi_{v}^{X}} .
\end{align*}
Here, $\hat{d}$ and $d_0$ are the electric dipole operator and the body-fixed dipole moment. The angular momentum dependence is described in the second line of the equation by the geometrical factors. $p$ denotes the spherical tensor component of the laser polarization, which can take the values of -1, 0, or +1 for $\sigma^{-}$, $\pi$, or $\sigma^{+}$ transitions respectively. The second term on the third line represents the transition dipole moment (TDM) evaluated at the Condon point $R_{\mathrm{C}}$. We adopt the ab initio results published in our previous work \cite{Yang2020} and replotted in Fig.~\ref{fig:SOCs} (c). At the calculated Condon point $R_{\mathrm{C}}=6.17\,\,a_{0}$, we estimate a TDM of 3.75 $ea_{0}$. $\bra{\psi_{v'}^{b}}\ket{\psi_{v}^{X}}$ is the wavefunction overlapping integral for the vibrational levels in $\mathrm{b}^{3}\Pi_{0}$ and $\mathrm{X}^{1}\Sigma^{+}$ states. Therefore, the transition dipole matrix element is estimated to be $2.6\times10^{-2}\,\, ea_{0}$ for $p=0$, corresponding to a partial linewidth of
\begin{equation}
	\Gamma=\frac{\omega^3\,d_{i,j}^{2}}{3\pi\epsilon_{0}\,\hbar c^3}=2\pi\times252\,\,\mathrm{Hz}
\end{equation}
to the $\ket{\mathrm{X}^{1}\Sigma^{+},v=0,J=0}$ state. 
\begin{figure*}[t]
	\centering
	\includegraphics[]{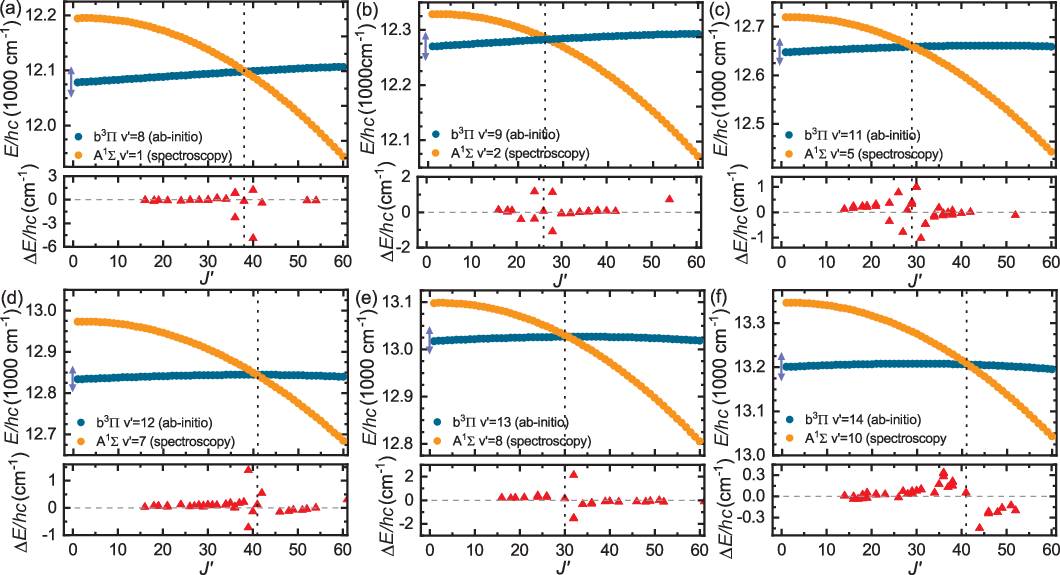}
	\caption{\label{fig:crossing}(a) - (f): Upper panels show the rotational energy progression of six vibrational states of the $\mathrm{A}^{1}\Sigma^{+}$ potential (orange circles) calculated using spectroscopically determined Dunham coefficients for $^{7}\mathrm{Li}^{39}\mathrm{K}$ \cite{Grochola2012, Tiemann2009} together with the rotational energies of a vibrational level in the $\mathrm{b}^{3}\Pi_{0}$ potential (blue circles), calculated using Dunham coefficients based on ab initio potentials \cite{Allouche}. Both $\mathrm{A}^{1}\Sigma^{+}$ and $\mathrm{b}^{3}\Pi_{0}$ transitions are derived from the same ground state $\ket{\mathrm{X}^{1}\Sigma^{+},v=0, J=J'-1}$ states. The blue curves are energetically shifted simultaneously as indicated by the light blue arrows such that they cross the orange curves at the center of the perturbed region. Lower panels show the error $\Delta E$ (red triangles) between experimentally observed transition energies for the $\mathrm{A}^{1}\Sigma^{+}$ potential and transitions calculated using Dunham coefficients. Observed transitions with $\Delta E$ higher than 0.1$\mathrm{cm}^{-1}$ are considered as perturbation caused by $\mathrm{b}^{3}\Pi_{0}$ rovibrational levels. The vertical dashed lines indicate the center for the perturbation region.}
\end{figure*}
\subsection{Prediction of resonance frequency}
With the ab initio potential curves \cite{Allouche} as shown Fig.~\ref{fig:PECs}, the energy difference between $\ket{\mathrm{X}^{1}\Sigma^{+},v=0,J=0}$ and $\ket{\mathrm{b}^{3}\Pi_{0},v'=0,J'=1}$ is calculated to be $hc\times10,490\,\mathrm{cm}^{-1}$. The largest uncertainty in the calculation is from the term energy $T_{\mathrm{e}}$ (referenced to the bottom of $\mathrm{X}^{1}\Sigma^{+}$) of the $\mathrm{b}^{3}\Pi_{0}$ potential. If we assume an uncertainty of $\geq2$\%, which is typical for ab initio calculations employing non-relativistic multireference configuration interaction with a moderate basis set, the corresponding uncertainty is $\geq hc\times210\,\mathrm{cm}^{-1}$ ($\geq h\times6\,$THz) for $\ket{\mathrm{b}^{3}\Pi_{0},v'=0,J'=1}$. In our experiment, the cycle time for taking one data point is approximately 3 min. Therefore a better prediction for the transition frequency is necessary to reduce the experimental effort for the spectroscopic search.

In the Supplement Information of the work by A. Grochola et al. \cite{Grochola2012} multiple perturbed spectral lines in the deeply bound region of the $\mathrm{A}^{1}\Sigma^{+}$ potential are included. These perturbations are caused by rovibrational states of the $\mathrm{b}^{3}\Pi_{0}$ potential. 
Most of the measured transitions can be described by a Dunham expansion\cite{Dunham1932}, some lines show clear signals of perturbation. This applies for lines with an energy deviation of $\Delta E >hc\times 0.1\,\mathrm{cm}^{-1}$ from the Dunham fit of the $\mathrm{A}^{1}\Sigma^{+}$ potential. In the lower panels of Fig.~\ref{fig:crossing}, $\Delta E/hc$ for some of the deeply-bound vibrational states of $\mathrm{A}^{1}\Sigma^{+}$ measured by Grochola \textit{et al.} \cite{Grochola2012} for $^{7}\mathrm{Li}^{39}\mathrm{K}$ with clear signals of perturbation are plotted as functions of rotational quantum number $J'$ for the excited states. In their work \cite{Grochola2012} these lines were discarded. However, we found that these perturbed lines may provide indications for locating the $\ket{\mathrm{b}^{3}\Pi_{0},v'=0}$. We calculated and plotted the energies for rotational levels of those vibrational states of the $\mathrm{A}^{1}\Sigma^{+}$ states, which are referenced against $\ket{\mathrm{X}^{1}\Sigma^{+},v=0, J = J'-1}$, using the experimentally determined Dunham coefficients \cite{Grochola2012,Tiemann2009} as shown with the orange curves in the upper panels of Fig.~\ref{fig:crossing}. The energy of a rovibrational state in an electronic potential can be expressed using the Dunham series
\begin{equation}
	E(v,J,\Omega) = \sum_{k,l} Y_{k,l}(v+\frac{1}{2})^{k}[J(J+1)-\Omega^{2}]^{l},
	\label{eqn:Dunham}
\end{equation}
where $Y_{k,l}$ are the Dunham coefficients precisely determined by fitting the spectroscopic data. For rovibrational states in the $\mathrm{A}^{1}\Sigma^{+}$ and $\mathrm{b}^{3}\Pi_{0}$ potentials the quantum number $\Omega$ for the projection of the total electronic angular momentum on the internuclear axis is zero.

For each perturbed feature in Fig.~\ref{fig:crossing}(a)-(f), a data point with low residual shift $\Delta E$ is selected, which we interpret to represent the center of the crossing. We use the $J'$ value of the selected point as assigned by Grochola et al.\cite{Grochola2012} to represent the center of perturbation. The selection rule for the perturbation of the rotational states due to spin-orbit coupling is $\Delta J=0$\cite{lefebvre-brion_field_2004}. Therefore, the curves describing the rotational energy progression of vibrational states in $\mathrm{b}^{3}\Pi_{0}$ potential should cross the curves for the vibrational levels of $\mathrm{A}^{1}\Sigma^{+}$ potential at the rotational quantum number $J'$s indicating the center of the perturbed regions. To calculate the energies for deeply-bound rovibrational states of the $\mathrm{b}^{3}\Pi_{0}$ potential we use the ab initio potential as shown in Fig.~\ref{fig:PECs}. The $\mathrm{b}^{3}\Pi_{0}$ potential curve is truncated to retain the lower half of the potential well, corresponding to energies below half of the potential depth $\mathcal{D}_{e}/2$. It is then re-parameterized by modifying the internuclear distance $R\rightarrow x=(R-R_{e})/R_{e}$, where $R_{e}$ is the equilibrium distance of the potential. The deeply-bound region of the curve is fitted with a polynomial function
$
f(x)=T_{e}+\sum_{i=2}^{10}\alpha_{i}x^{i},
$
where $T_{e}$ is the energy offset for the potential bottom. The Dunham coefficients $Y_{k,l}$ of $\mathrm{b}^{3}\Pi_{0}$ can be expressed by combinations of the fitted parameters $\alpha_{i}$ and the equilibrium rotational constant $B_{e} = \hbar^2/(2\mu R_{e}^2)$ \cite{Dunham1932}. $\mu$ is the reduced mass of $^{7}\mathrm{Li}^{39}\mathrm{K}$ isotopologue. The above process is repeated for the ab initio $\mathrm{X}^{1}\Sigma^{+}$ potential curve for extracting the Dunham coefficients. The calculated energies of $\ket{\mathrm{b}^{3}\Pi_{0},v',J'}$ using the fitted Dunham coefficients, which are referenced against $\ket{\mathrm{X}^{1}\Sigma^{+},v=0,J=J'-1}$, are shown by the blue curves in the upper panels of Fig.~\ref{fig:crossing}. As a second step to achieve crossings in the perturbation region we shift $T_{e}$ and hence shifting energies of the rovibrational states of $\mathrm{b}^{3}\Pi_{0}$ simultaneously. At a shift of -\,8\,$\mathrm{cm}^{-1}$, good agreement is found between the rotational quantum number $J'$, at which the two curves intersect, and the center of the perturbation. This is quantitatively supported by minimizing the sum, over all six sub-figures, of the squared error between the $\mathrm{A}^{1}\Sigma^{+}$-$\mathrm{b}^{3}\Pi_{0}$ rovibrational energy differences at $J'$ and a variable energy offset. This means that the minimum of the ab-initio adiabatic potential for $\mathrm{b}^{3}\Pi_{0}$ is approximately 8\,$\mathrm{cm}^{-1}$ too high, independent of the choice of isotopologue. 

In the following we apply the shift to all predicted lines for $^{6}\mathrm{Li}^{40}\mathrm{K}$ isotopologue and obtain a transition wavenumber for $\ket{\mathrm{X}^{1}\Sigma^{+},v=0,J=0}\rightarrow\ket{\mathrm{b}^{3}\Pi_{0},v'=0,J'=1}$ of 10,481.9 $\mathrm{cm}^{-1}$, which corresponds to the laser frequency of 314,239 GHz. Here, we give a conservative uncertainty range of $\pm5\,\mathrm{cm^{-1}}$ for the prediction of the $v'=0$ state. Apart from $T_{e}$, deviations in the potential shape will also affect the prediction for the transition frequencies. However, the uncertainty due to the leading Dunham term $\omega_{e}$, which describes the harmonic vibration frequency of the potential, is expected to be only $1\,\mathrm{cm^{-1}}$ assuming a 1\% error for the ab initio calculations. Its contribution to the total uncertainty in the energy of $\ket{\mathrm{b}^{3}\Pi_{0},v=0,J=1}$ is therefore almost negligible. Based on the above empirical analysis the uncertainty is now significantly reduced from approximately 3 THz to 300 GHz, corresponding to $\pm5\,\mathrm{cm^{-1}}$. Indeed, as will be shown later in Sec.~\ref{sec:measurement}, the transition was found within the interval.

\subsection{Laser Setup}
For achieving high laser power and frequency tunability, we built an external-cavity diode laser (ECDL) and use it to seed a tapered amplifier (TA) for power amplification. The laser diode has an anti-reflection coating to allow for wavelength tuning from 940 nm to 980 nm. The free-running short-term linewidth of the ECDL is around 2 MHz. The TA is seeded with approximately 20 mW laser power and provides an output power up to 700 mW. In the experiment, the wavelength of the laser is monitored by a wavemeter (WS-7, High Finesse) and adjusted in 500 MHz steps when searching for the resonance. In the experiment, we focused around 500 mW of laser power on the molecules to a $1/e^2$ Gaussian-beam waist of 80 $\mu\mathrm{m}$, which allows for an estimated laser Rabi frequency of $\Omega = 2\pi\times 65 \,\,\mathrm{MHz}$.

\begin{figure}[h]
	\centering
	\includegraphics[width=8.8cm]{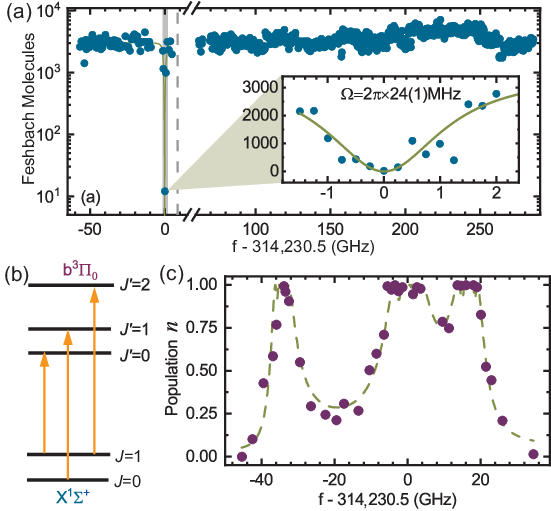}
	\caption{\label{fig:loss} Spectroscopy measurement for $\mathrm{b}^{3}\Pi_{0}$. (a) The number of Feshbach molecules is recorded (blue circles) while scanning the frequency of spectroscopy laser around $\nu_{0} = 314230.5 \rm{GHz}$. The inset shows a scan of laser frequency around $\nu_{0}$ with finer steps. The solid green curve is a fit to the data, from which we extract an Rabi frequency $\Omega$ of 24 MHz. Details of the fit is explained in the text. (b) Illustration of rotational level scheme for the ground and target excited state. (c) Population of molecules is recorded after a microwave $\pi$ pulse (purple circles), which drives the transition $\ket{J=0,m_{j}=0}\rightarrow\ket{J=1,m_{j}=-1}$. The spectroscopy laser is switched on during the process. The three peaks are caused by the $\ket{J=0}\rightarrow\ket{J'=1}$, $\ket{J=1}\rightarrow\ket{J'=0,2}$ transitions. Green dashed curves are fit to the data.}
\end{figure}

\subsection{Measurement}
\label{sec:measurement}
In each measurement, $^{6}\mathrm{Li}$-$^{40}\mathrm{K}$ Feshbach molecules are first transferred to the ground state $\ket{\mathrm{X}^{1}\Sigma^{+},v=0,J=0}$ and transferred back to the Feshbach state using STIRAP for detection. During this process, the spectroscopy beam at 954nm is switched on to introduce an AC Stark shift. When the spectroscopy laser scans in the far-off-resonant range, the three-level STIRAP system experiences an almost constant energy shift, which is considered as a fixed background and compensated by shifting the frequency of the STIRAP lasers. The half-width at half-maximum (HWHM) of the STIRAP two-photon line profile is measured to be 112(10) kHz. For the given laser intensity and transition strength estimated in Section~\ref{sec:transitionstrength}, we will only be able to observe significant molecular number losses when the spectroscopy laser is within a range of $\pm$10 GHz. The losses are induced by the AC Stark shift of the STIRAP two-photon resonance for equal or larger than one HWHM. As the uncertainty in $\langle\hat{d}\rangle$ is dominated by $c_{1}$, which may be overestimated by up to one order of magnitude, the scanning step of the laser frequency is chosen to be 1 GHz to avoid missing the transition. The measured molecular loss spectrum for a frequency range from 314,173.5 GHz to 314,516.0 GHz is shown below in Fig.~\ref{fig:loss} (a). A narrow resonance is found to be centered at 314,230.5(5) GHz. The uncertainty in the line center is due to the instability of our spectroscopy laser. As the bottom of $\mathrm{b}^{3}\Pi_{0}$ potential is much deeper than any other excited-state potentials by more than 30 THz, the measured transition resonance can only be explained by the transition from $\ket{\mathrm{X}^{1}\Sigma^{+},v=0,J=0}$ to $\ket{\mathrm{b}^{3}\Pi_{0},v'=0,J'=1}$. A scan with finer steps around the transition resonance is shown in the inset of Fig.~\ref{fig:loss} (a). The loss feature is modeled by a Lorentzian distribution $w^2/\pi(\delta(x,\Omega)^2+w^2)$ with $\delta(x,\Omega)$ being the AC Stark shift as a function of laser detuning $x=\mathrm{f}-\nu_0$ and the Rabi frequency $\Omega$ \cite{Autler1955}. $w$ is the HWHM of the STIRAP two-photon linewidth of our experiment, which was measured to be 112(7) kHz. By fitting the decay curves, we extract a Rabi frequency $\Omega$ of $24(1)\,\,\mathrm{MHz}$. This indicates a much smaller transition dipole matrix element for the $\mathrm{X}\rightarrow b$ transition of $9.6\times10^{-3}e\,a_{0}$.

Further, we probe the rotational structure of the excited state and locate the energy for the rovibrational ground state of $\mathrm{b}^{3}\Pi_0$ potential. We use a microwave (MW) signal at 17.482 GHz to drive the Rabi oscillation between the ground state $J=0$ and the first excited state $J=1$ of the $\ket{\mathrm{X}^{1}\Sigma^{+},v=0}$ for a duration of 400 $\mu\mathrm{s}$, corresponding to half of a period (or $\pi$ pulse). The spectroscopy laser is switched on for a duration that covers the MW pulse. When the laser frequency is far off-resonance from the optical transitions to the rotational manifold of the $\ket{\mathrm{b}^{3}\Pi,v'=0}$ as illustrated in Fig.~\ref{fig:loss} (b), the induced AC Stark shift can be treated as a fixed background and is compensated by shifting the MW frequency. While scanning laser frequency, the residual molecule population $n$ in $\ket{X^1\Sigma^{+},v=0,J=0}$ is recorded after each MW-$\pi$ pulse and the result is shown in Fig.~\ref{fig:loss} (c). At large laser detunings, the MW pulse drives the resonant Rabi $\pi$-pulse and transfers the full population to $\ket{X^1\Sigma^{+},v=0,J=1}$, leading to depletion of $n$. Three peaks are observed while scanning the laser frequency close to the three transitions $J=0\rightarrow J'=1$ \& $J=1\rightarrow J'=0,2$, which locate at the detunings of $h\,\Delta_{i}=$0, $-2 B_{0}^{X}- 2 B_{0}^{b}$, and $4 B_{0}^{b}-2 B_{0}^{X}$. $B_{0}^{X}$ and $B_{0}^{b}$ are the rotational constants for $\mathrm{X}^{1}\Sigma^{+}$ \& $\mathrm{b}^{3}\Pi_{0}$ vibrational ground states. The increase in population at these laser detunings can be modeled by a Rabi oscillation processes at a detuning due to laser-induced AC Stark shift $\delta_{\rm{tot.}}$, which is contributed by the sum of the three transitions $\delta_{\mathrm{tot.}} = \sum_{i}\delta(x,\Delta_{i})$. $x=\mathrm{f}-\nu_{0}$ is the laser detuning. $B_{0}^{X}/h = 8.742(3)\,\, \mathrm{GHz}$ was previously determined experimentally \cite{Yang2020}. By fitting the data using the model described above, we obtained $B_{0}^{b}/h = 8.576(44) \,\,\mathrm{GHz}$. Here, we do not consider the effect of the centrifugal distortion $D_{e}$ on the position of the resonances. From the Dunham coefficients calculated using the ab initio curves we obtain a $D_{e}/h$ of 73.5 kHz, which is four orders of magnitude smaller than the data resolution. The obtained rotational constant $B_{0}^{b}$ is 9\% smaller than our calculated value based on the equilibrium distance $R_{e} = 6.086 \, \mathrm{a}_{0}$ of the ab initio potential\cite{Allouche}. This suggests that the calculations underestimated $R_{e}^{b}$ by approximately 4\% for the published ab initio potentials that we have investigated\cite{Allouche,Rousseau1999,Miadowicz2013,Musial2016}. A similar trend was observed in the de-perturbation analysis of the $\mathrm{A}^{1}\Sigma^+-\mathrm{b}^{3}\Pi_{0}$ state complex of LiCs\cite{Grochola2014}. For LiCs, the obtained molecular constants for the $\mathrm{A}^{1}\Sigma^+$ state agrees well with ab initio calculations\cite{Elkork2009}, while the agreement is significantly poorer for the $\mathrm{b}^{3}\Pi_0$ state. Moreover, the energy of $\ket{\mathrm{b}^{3}\Pi_{0},v'=0,J'=0}$ relative to $\ket{\mathrm{X}^{1}\Sigma^{+},v=0,J=0}$ is determined to be $E^{b}_{0}= h\times314,213.3(5)$ GHz, corresponding to the wavenumber of 10,481.03(2) $\mathrm{cm}^{-1}$. As the binding energy for $\ket{\mathrm{X}^{1}\Sigma^{+},v=0,J=0}$ has been precisely measured by our group using two-photon spectroscopy to be 6,104.18081(1) $\mathrm{cm}^{-1}$ \cite{Yang2020}, the binding energy for $\ket{\mathrm{b}^{3}\Pi_{0},v'=0,J'=0}$ is determined to be 8,666.05(2) $\mathrm{cm}^{-1}$ relative to the $^{40}\mathrm{K}$ D2 asymptote.

\section{Conclusion}
As a first step toward achieving magic conditions for the rotational coherence of ultracold $^{6}\mathrm{Li}^{40}\mathrm{K}$ ground state molecules, we performed molecular spectroscopy to search for the weak transition from the absolute ground state of $\mathrm{X}^{1}\Sigma^{+}$ potential to the vibrational ground state of $\mathrm{b}^{3}\Pi_{0}$ excited state potential. The prediction for the transition frequency is assisted by analyzing the available spectroscopy data on perturbed transitions of deeply bound state in the $\mathrm{A}^{1}\Sigma^{+}$ potential \cite{Grochola2012}. We first determined the transition frequency to be 314,230.5(5) GHz. By using a microwave field to couple the rotational ground and first excited state $J=0,1$ of $\ket{\mathrm{X}^{1}\Sigma^{+},v=0}$, we spectroscopically resolved the rotational structure ($J'=0,1,2$) of the $\ket{\mathrm{b}^{3}\Pi_{0},v'=0}$ and extracted the rotational constant $B_{0} = h\times8.576(44)$ GHz. By replacing the diode laser setup with a high-power fiber-laser system frequency stabilized to an ultra-stable high-finesse cavity, we can achieve optical lattices operating at magic conditions to obtain long coherence times for ultracold $^{6}\mathrm{Li}^{40}\mathrm{K}$ molecules interacting via strong dipole-dipole interactions. Our work also benefits the study of fermionic LiK isotopologues in their applications in quantum simulation and quantum computation in optical tweezer arrays.

\section*{Author contributions}
A.Y. conceived the idea, provided predictions/corrections for the transition frequency, participated in the measurement, found the transition and wrote the manuscript; X.N. built the laser setup, predicted the transition strengths, conducted the measurement and found the transition; H.L.Y., Y.L., V.A. and C.H. participated in the measurement; J.K and S.K. conducted calculations for the SOC coupling, provided theoretical guidance, and helped write the manuscript; K.D. is the supervisor of this experimental work and provided guidance to the team; All authors have reviewed the manuscript.

\section*{Conflicts of interest}

There are no conflicts to declare.

\section*{Data availability}
Data supporting this article, including the calculations of Franck–Condon factors, spin–orbit coupling strengths, predicted rovibrational transition frequencies, and spectroscopy measurements are available on Zenodo at https://doi.org/10.5281/zenodo.17386208.

\section*{Acknowledgements}
The authors would like to thank Sofia Botsi and Sunil Kumar for their earlier contributions in the lab. The research work at Centre for Quantum Technologies and National University of Singapore is supported by the National Research Foundation, Singapore and A*STAR under its CQT Bridging Grant. The work at Temple University was funded by the AFOSR, Grant No. FA9550-21-1-0153, the National Science
Foundation, Grant No. PHY-2409425, and the Gordon and Betty Moore Foundation, Grant Id. GBMF12330




\bibliography{references} 
\end{document}